\documentclass[twocolumn,amsmath,amssymb,aps,superscriptaddress]{revtex4}

\usepackage[dvips]{graphicx}
\usepackage{verbatim}

\newcounter{saveeqn}
\newcommand{\alpheqn}{\setcounter{saveeqn}{\value{equation}}%
  \stepcounter{saveeqn}\setcounter{equation}{0}%
  \renewcommand{\theequation}
    {\mbox{\arabic{saveeqn}-\alph{equation}}}}
\newcommand{\reseteqn}{\setcounter{equation}{\value{saveeqn}}%
  \renewcommand{\theequation}{\arabic{equation}}}

\begin{document}   


\title{Core excitation in O$_3$ localized to one of two symmetry-equivalent chemical bonds\\- molecular alignment through vibronic coupling}

\author{K. Wiesner\footnote{Current address: Center for Computational Science and Engineering, University of California, Davis, CA 95616, USA}}
\affiliation{Department of Physics, Uppsala University, Box 530, 751 21 
Uppsala, Sweden}
\author{A. Naves de Brito\footnote{On leave from: Institute of Physics, University of Brasilia, Brazil}}
\affiliation{Laborat\'{o}rio Nacional de Luz S\'{\i}ncrotron, Box 
o6192 CEP:13084-971 Campinas - SP Brazil}
\author{S. L. Sorensen}
\affiliation{Department of Synchrotron Radiation Research, Institute of 
Physics, University of Lund, Box 118, 221 00 Lund, Sweden}
\author{N. Kosugi}
\affiliation{UVSOR, Institute for Molecular Science, Myodaiji, Okazaki 444-8585, Japan}
\author{O. Bj\"{o}rneholm}
\affiliation{Department of Physics, Uppsala University, Box 530, 751 21 
Uppsala, Sweden}

\begin{abstract}
Core excitation from terminal oxygen O$_T$ in O$_3$ is shown to be an excitation from a localized core orbital to a localized valence orbital. The valence orbital is localized to one of the two equivalent chemical bonds. We experimentally demonstrate this with the Auger Doppler effect which is observable when O$_3$ is core-excited to the highly dissociative O$_{ T}$1s$^{-1}$7a$_1^1$ state. Auger electrons emitted from the atomic oxygen fragment  carry information about the molecular orientation relative to the  electromagnetic field vector at the moment of excitation. The data together with analytical functions for the electron-peak profiles give clear evidence that the preferred molecular orientation for excitation only depends on the orientation of one bond, not on the total molecular orientation. The localization of the valence orbital "7a$_1$"  is caused by mixing of the valence orbital "5b$_2$" through vibronic coupling of anti-symmetric stretching mode with b$_2$-symmetry. To the best of our knowledge, it is the first discussion of the localization of a core excitation of O$_3$. This result explains the success of the widely used assumption of localized core excitation in adsorbates and large molecules.
\end{abstract}

\maketitle
\section{Introduction}

A successful model of molecular valence electronic structure is the formation of delocalized molecular valence orbitals out of localized atomic valence orbitals. Core-to-valence excitation in small molecules is therefore treated as a delocalized process. On the other hand core-valence interaction in large molecules is often assumed to be limited to the site of the core-excited atom. This is the case even for  molecules with several symmetry-equivalent atoms, thus allowing for a symmetry break. When is it reasonable to treat a core-valence excitation as localized? We present O$_3$, which despite being a small molecule, exhibits core-excitation localized to one of the two symmetry-equivalent bonds O$_C$-O$_T$ (O$_C$ and O$_T$ are the center and terminal atoms). We attribute this effect to a vibronic coupling between two core excited states involving the valence orbitals 7a$_1$ and 5b$_2$. This causes an energetically favorable symmetry lowering and allows for the localization.

Core excited O$_3$ has been studied both with ion and electron spectroscopy \cite{gejo:99,deBrito:00,stranges:01,rosenqvist:01,wiesner:03}. The core-excited state O$_{ T}$1s$^{-1}$7a$_1^1$ has been shown to be ultra-fast dissociating \cite{rosenqvist:01}. Ultra-fast dissociating  molecules dissociate on the same time scale as the electronic Auger decay, which is in the femtosecond range. We take advantage of this characteristic to obtain information about the preferred molecular orientation relative to the electric light vector (e-vector) during core excitation. The Resonant Auger Electron (RAE) signal from the O$_3$ molecules that dissociate before they undergo RAE decay ($\approx 10~\%$) is an atomic oxygen RAE signal. The atomic Auger electrons obtain additional momentum from the emitting fragment as a consequence of the  preceding dissociation. This additional momentum is detected as a shift in energy in the RAE spectrum and can be used to determine the dissociation direction of the fragment. Since we study a two-body dissociation this direction gives full information about the molecular orientation before dissociation. The molecular orientation before dissociation is to a very good approximation identical to the orientation during excitation since rotation is two orders of magnitude slower than the Auger decay. Hence, the RAE spectrum from the fragment contains information about the molecular orientation at the time of excitation.

In this paper we present calculatations of the molecular orientation during core excitation and find that one chemical bond is preferably parallel to the electric-field vector. This represents a symmetry break which we explain with the well-known concept of vibronic coupling.

\section{Experiment}

The experimental data shown in Fig.~\ref{fig:exp} were published in \cite{rosenqvist:01}. The core-excited state  investigated was the O$_{ T}$1s$^{-1}$7a$_1^1$ at 535.85~eV excitation energy. In literature this core resonance is often designated as ``$\sigma^{*}$'' resonance.
 The experiments were performed at the undulator beam line I411 \cite{bassler:01} at the MAX II storage ring of the Swedish National Synchrotron Laboratory in Lund, Sweden. 
 The RAE spectra were measured with a photon energy resolution of 120~meV and an electron spectrometer resolution of 140(10)~ meV. The ozone sample was generated in a commercial ozone generator (Ozone Technology, Sweden) by discharging oxygen (O$_{2}$) of industrial quality in an electric field. The O$_{2}$/O$_3$ mixture was then distilled to a purity of $\approx 99~\%$.  For further experimental details, see the original paper \cite{rosenqvist:01}.

\section{Functions for Electron-Peak Profiles}
\label{sec:functions}

\begin{figure}
\includegraphics[scale=.4]{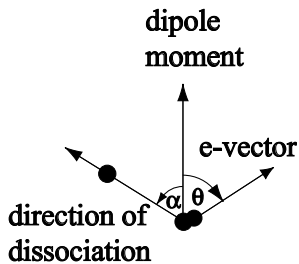}
\caption{\label{fig:CS}Definition of the angles $\theta$ and $\alpha$.}
\end{figure}

\begin{figure*}
\includegraphics[scale=.5]{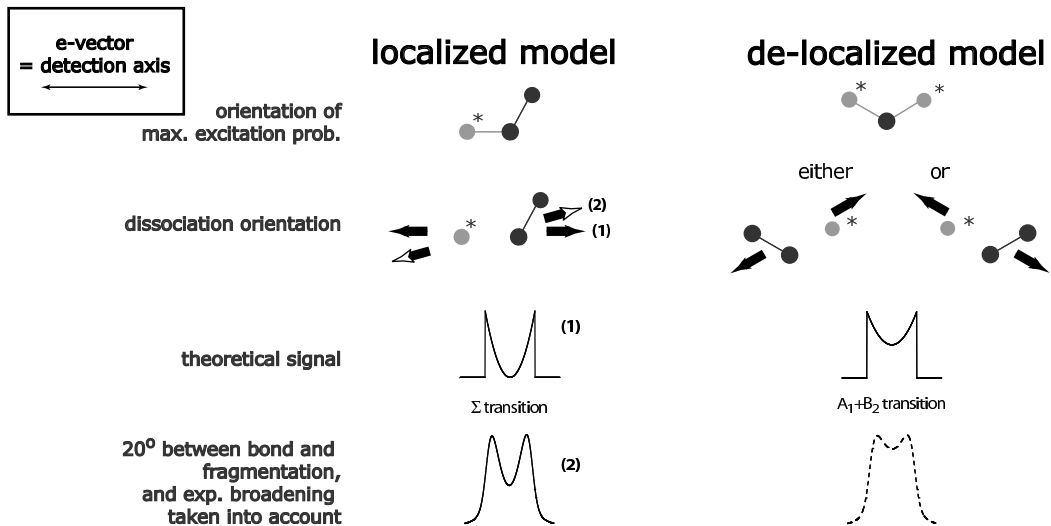}
\caption{\label{fig:process}The excitation and dissociation process according to the model for localized (left) and delocalized excitation (right), respectively. The core-excited atoms (*) are grey-shaded. The profiles at the bottom are added to the measured spectrum in Fig.~\ref{fig:exp}.}
\end{figure*}

 The O$_3$ molecule belongs to the symmetry group $C_{2\nu}$ with four irreducible representations for electronic orbitals: a$_1$, a$_2$, b$_1$ and b$_2$. To be  consistent with the nomenclature in the original experimental paper we define the molecule lying in the $yz$-plane. The O$_{ T}$1s atomic core orbitals combine {\it gerade} to form the molecular orbital (MO) 2a$_1$ and {\it ungerade} to form 1b$_2$. The transition to the unoccupied 7a$_1$ MO is dipole-allowed from both the 2a$_1$ and the 1b$_2$ core MOs. Thus, the transition to the O$_{ T}$1s$^{-1}$7a$_1^1$ core-excited state is actually two transitions from the two nearly-degenerate O$_{ T}$ core MOs. The symmetry of the transition is $A_1$ and $B_2$, respectively. In the following it is assumed and likely to be correct that the transition probability for the two is the same.\\
To investigate how localized and delocalized descriptions of core excitation in O$_3$ lead to experimentally observable differences, we derived analytical functions for peak profiles of electron spectra for different transition symmetries. The premises for the two different models, the localized and the delocalized, are summarized in Fig.~\ref{fig:process}.

We will start with the derivation for the case of a diatomic molecule which is the simplest case and the result will be of use later. We simplify the derivation by aligning the detection axis and the e-vector (for the angle definitions, see Fig.~\ref{fig:CS}). The angular distribution for photoexcitation, $f(\theta)$, about the polarization of the exciting radiation (e-vector) (for a one-photon excitation in the dipole-approximation) is given by \cite{fano:72}:

\begin{equation}
f(\theta)=\frac{1}{4\pi}[1+\beta P_2(cos\theta)],
\end{equation}

where $\theta$ is the angle between the molecular symmetry axis and the e-vector, $\beta$ is the anisotropy parameter and $P_2(cos\theta)=\frac{1}{2} (3cos^2\theta-1)$ is the Legendre polynomial of order $2$. The spatial anisotropy parameter $\beta$ can range from $+2$ for a pure parallel transition to $-1$ for a pure perpendicular transition. In the present experiment we are sensitive only to electrons from ultra-fast dissociated molecules. The timescale of autoionization for those electrons is a few femtoseconds, which is about two orders of magnitude faster than molecular rotation. Thus measurement of $\beta$ provides information about the symmetry of the transition.

In order to obtain a theoretical electron profile for a two-body dissociation following a transition with a certain $\beta$ parameter, we need to integrate the angular distribution over all angles, since we describe a sample of randomly oriented molecules. The extra fragment momentum from dissociation is transferred to the electron. The measurement determines the projection $cos\,\theta$ of the momentum along the detection axis. With the e-vector and the detection axis aligned, the integral over an infinitesimal interval $\delta\:\theta$ becomes:

\begin{eqnarray}
I(cos\,\theta) & = & \frac{1}{\delta}\,\int\limits^{\theta}_{\theta + \delta}\int\limits^{2\pi}_{0} f(\theta^{\prime}) \,\mathrm{d}\theta^{\prime}\mathrm{d}\varphi\\
 \delta \rightarrow 0\\
\label{eqn:I_theta}
& = & \frac{1}{2}(1-\frac{\beta}{2})+\frac{3}{4}\beta \,cos^2\theta
\end{eqnarray}

The kinetic energy released in the dissociation will broaden the profile. When scaling the profile to electron kinetic energy, the maximum width of the profile is: 

\begin{equation}
\label{eqn:delta}
\Delta(cos\,\theta)_{max}\,=\,2\frac{m_f}{m_e}E_f
\end{equation}

with the kinetic energy of the fragment $E_f$, the electronic mass $m_e$ and the fragment mass $m_f$. 

We write out the explicit form of the function for a $\Sigma$ transition ($\beta = 2$):
\begin{eqnarray}
\label{eqn:s}
I(cos\theta)_{\Sigma} &=& \frac{3}{2} cos^2\,\theta.
\end{eqnarray}

This is different from the electron-peak profiles derived in \cite{bjorneholm:01}, since this reference uses an invalid geometric factor. The electron profile for the $\Sigma$ transition is shown in Fig.~\ref{fig:process} on the left.

For the details of the derivation of the electron-peak profiles for bent molecules of $C_{2\nu}$ symmetry we refer to Appendix A. Here we only give the final result for the sum of the two transitions $A_1$ and $B_2$ relevant to our experiment:

\begin{eqnarray}
\label{eqn:a1b2}
I_{A_1+B_2} &=&  \frac{3}{8} (1+cos^{2}\theta).
\end{eqnarray}

 The profile is shown in Fig.~\ref{fig:process} on the right. Eqn.~\ref{eqn:s} for a profile of a diatomic molecule can be used for describing excitation localized to one bond in a polyatomic molecule. Eqn.~\ref{eqn:a1b2} represents excitation delocalized over all bonds.\\

\section{Results}

\begin{figure}
\includegraphics[scale=.5]{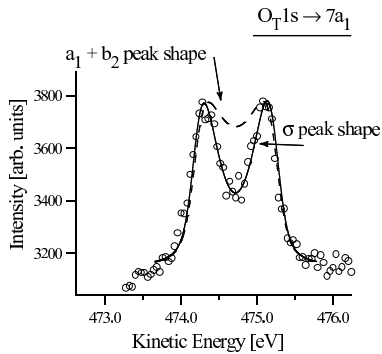}
 \caption{\label{fig:exp}RAE spectrum of O$_3$ (circles) excited to O$_{ T}$1s$^{-1}$7a$_1^1$. The solid and dotted lines are the profiles from Fig.~\ref{fig:process} for $\Sigma$ transition and A$_1+$B$_2$ transition, respectively.}
 \end{figure}

We now compare the experimental data with the analytically derived profiles for localized and delocalized excitation, Eqns.~\ref{eqn:s} and \ref{eqn:a1b2}. A fragment Auger electron peak following O$_{ T}$1s$^{-1}$7a$_1^1$ excitation \cite{rosenqvist:01} is shown in Fig.~\ref{fig:exp}. The analysis of the other fragment peaks in Ref.~\cite{rosenqvist:01} is analogous and leads to the same conclusions. The electron signal is double-peaked due to the Doppler effect described in \cite{rosenqvist:01}. The profile that should apply if a delocalized description of this transition was valid is the $A_1$+$B_2$ profile, Fig.~\ref{fig:process} (right) and Eqn.~\ref{eqn:a1b2}. We convoluted the profile with a Gaussian and a Lorentzian function. The Gaussian function with a FWHM of 140~meV accounts for the limited experimental resolution. The Lorentzian function with FWHM of 149~meV accounts for the lifetime broadening, where we assumed the same core hole lifetime as that of O$_2$ \cite{coreno:99}. The convoluted profile is shown as a dashed line in Figs.~\ref{fig:process} (right) and \ref{fig:exp}. The solid line in Fig.~\ref{fig:exp} is the likewise convoluted profile for $\Sigma$ transition, also shown in  Fig.~\ref{fig:process} (left). We assume an angle of 20$^{\circ}$ between fragmentation and symmetry axis. This angle accounts for momentum conservation. In addition an angle of 7$^{\circ}$ between the orbital and the bond is taken into account, as calculated by Oji {\it et al.} \cite{oji:note}. The only free parameter in both functions is $\Delta(cos\theta)_{max}$, which depends on the kinetic energy released by dissociation, see Eqn.~\ref{eqn:delta}. 


\section{Discussion}

\begin{figure}
\includegraphics[scale=.5]{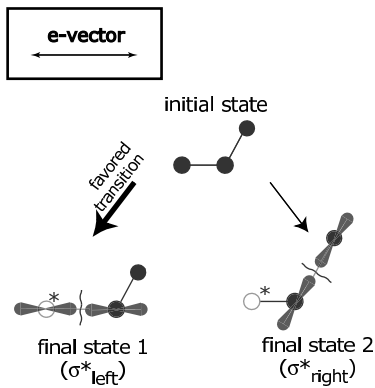}
 \caption{\label{fig:finalstate}Illustration of valence orbital localization after excitation. The prefered molecular orientation is with the bond including the core-excited atom (*) parallel to the e-vector. The atomic 2p orbital is indicated, illustrating the localized $\sigma^*$ orbital.}
 \end{figure}

It is clear from Fig.~\ref{fig:exp} that the sum of electron-peak profiles for $A_1$ and $B_2$ transition (dashed line) does not reproduce the experiment. The  profile for a $\Sigma$ transition (solid line), on the other hand, agrees very well with the experiment. The two profiles represent two different models for excitation, see Sec.~\ref{sec:functions} and Fig.~\ref{fig:process}. The $A_1$+$B_2$ profile represents a transition delocalized over the molecule, while the $\Sigma$ profile represents a transition localized to one bond. It is obviously the latter case we observe here, a transition localized to one bond. The excitation probability for each bond is given as a probability, determined by its orientation relative to the light vector and is independent on the orientation of the other bond. In other words the molecular symmetry is broken. This result is schematically summarized in Fig.~\ref{fig:finalstate}. A discussion of the localization follows in the next paragraph.

An explanation for the observed localization can be found in the mechanism  of {\it vibronic coupling} \cite{domcke:77,cederbaum:95}. Whenever two states of the same total symmetry (electronic $+$ vibronic) are close in energy, they can couple to each other. Thus states of different electronic symmetry can couple via suitable vibrational modes. If they couple via an asymmetric mode of b$_2$-symmetry the molecular symmetry is lowered. This is what happens in the given core excitation in O$_3$. The two core-excited states 1s$_T^{-1}$7a$_1^1$ and 1s$_T^{-1}$5b$_2^1$ are close in energy and are not experimentally resolved. Coupling between those states opens up for an asymmetric vibrational mode that lowers the symmetry and localizes the two valence orbitals to opposite bonds. In the lower symmetry they  form two $\sigma^*$ orbitals: 7a$_1\pm$5b$_2\,=\,\sigma^*_{left/right}$ (normalization factors neglected), each localized to the respective bond. Thus, instead of an excitation of a localized core electron to a delocalized valence orbital we obtain an excitation to a localized valence orbital. In other words, through vibronic coupling the four delocalized excitations, expressed in $C_{2\nu}$ symmetry: 
\alpheqn
\begin{eqnarray}
1s_{ T-left}  \rightarrow  7a_1 & \;\pm\; & 1s_{ T-right} \rightarrow 7a_1 \\
1s_{ T-left} \rightarrow 5b_2 & \;\pm\; & 1s_{ T-right} \rightarrow 5b_2 
\end{eqnarray}
\reseteqn
are turned into two localized excitations (final state 1 in Fig.~\ref{fig:finalstate})\footnote{We express the excitations in the point group of diatomic molecules (C$_{\infty \nu}$) to emphazise the localized character, although the actual molecular symmetry after excitation is C$_S$.}:

\alpheqn
\begin{eqnarray}
1s_{ T-left} & \rightarrow & \sigma^*_{left}, \\
1s_{ T-right} & \rightarrow & \sigma^*_{right}, 
\end{eqnarray}
\reseteqn
assuming $1s_{ T-left} \, \rightarrow \, \sigma^*_{right}$ (final state 2 in Fig.~\ref{fig:finalstate}) and $1s_{ T-right} \, \rightarrow \, \sigma^*_{left}$ is weak \cite{deBrito:00}, as shown in Fig.~\ref{fig:finalstate}.

\section{Conclusion}

Core excitation in O$_3$ is found to be localized to one chemical bond. Analytically derived functions for the electron-peak profiles were used to identify the excitation geometry of the molecule revealing the fact that the  O$_{ T}$1s$^{-1}$7a$_1^1$ core-excited state is preferably created with one O$_{ T}^*$-O$_{ C}$ chemical bond aligned to the e-vector ($^*$ indicating the core-excited atom). The model assumed to date for small molecules, delocalizing the excitation over the two symmetry-equivalent bonds, does not fit the data. We explain the localization of the excitation with a symmetry breaking of the valence orbital caused by vibronic coupling. Our result provides an experimental verification for the widely used assumption of core excitation localized to one of several symmetry-equivalent bonds.

\acknowledgments

K.~J.~B{\o}rve, A.~Lindgren and J.~{\AA}berg are gratefully acknowledged for helpful discussion. This work was supported by the Foundation for Strategic Research (SSF), the Swedish Research Council (VR), the G\"{o}ran Gustavsson foundation as well as by the Swedish Foundation for International Cooperation in Research and Higher Education (STINT). ANB would like to thank CNPq-Brazil and FAPESP-Brazil for financial support.\\

\appendix{Appendix A}
\vspace{10pt}
\par We derive the explicit electron-peak profiles for $A_1$ and $B_2$ transitions of bent molecules of $C_{2\nu}$ symmetry. The main difference compared to the diatomic-molecule case is that the axis of fragmentation and the symmetry axis are not aligned. In other words the angular distribution for photoexcitation depends on the angle between the e-vector and the symmetry axis, whereas the measured coordinate depends on the angle between the e-vector (which is equal to the detection axis) and the fragmentation axis (see Fig.~\ref{fig:CS}). We therefore need to take into account the angle $\alpha$ between symmetry axis and axis of fragmentation. The way this is done is by rotating the reference frame of the transition dipole into the reference frame of the axis of fragmentation. 

 The angular distribution for photoexcitation, $f(\theta)$, about the polarization of the exciting radiation (e-vector) for an excitation of $A_1$ symmetry is given by $f(\theta)=\frac{3}{4\pi}cos^2\,\theta$. The transition dipole is aligned with the symmetry axis of the molecule. We need to integrate the angular distribution over all angles, but in the frame of dissociation. Thus, we have to adopt a new coordinate system which is rotated by the angle $\alpha$ with respect to the old one, see Fig.~\ref{fig:CS}. The new coordinate system is obtained through a Eulare rotation and $\tilde f(\theta)$ reads:

\begin{equation}
\tilde f(\theta)_{A_1}=\frac{1}{4\pi}[sin^2\alpha - (1-cos^2\alpha) cos^2\theta]
\end{equation}

The transition dipole of a $B_2$ transition is perpendicular to the symmetry axis and lies in the molecular plane. The angular distribution is given by $f(\theta)=\frac{3}{4\pi}sin^2\,\theta sin^2\varphi$. Transformed into the rotated coordinate system it reads:

\begin{equation}
\tilde f(\theta)_{B_2}=\frac{1}{4\pi}[cos^2\alpha - (1-sin^2\alpha) cos^2\theta]
\end{equation}

The integration is performed as in the diatomic case:

\begin{eqnarray}
I(cos\theta) & = & \frac{1}{\delta}\,\int\limits^{\theta}_{\theta + \delta}\int\limits^{2\pi}_{0} f(\theta^{\prime}) \,\mathrm{d}\theta^{\prime}\mathrm{d}\varphi\\
 \delta \rightarrow 0\\
I(cos\theta)_{A_1} & = & \frac{3}{4} (2 - 3\, sin^2\,\alpha) \:cos^2\theta +\frac{3}{4}\, sin^2\,\alpha \\
I(cos\theta)_{B_2} & = & \frac{3}{4} (2 - 3\, cos^2\,\alpha) \:cos^2\theta +\frac{3}{4}\, cos^2\,\alpha 
\end{eqnarray}

Since the core excitation O$_{ T}$1s$^{-1}$7a$_1^1$ consists of two overlapping transitions of symmetry $A_1$ and $B_2$ we need to sum the intensity profiles. The final intensity profile becomes:

\begin{eqnarray}
I(cos\theta)_{A_1} + I(cos\theta)_{B_2} =  \frac{3}{8} (1+cos^{2}\theta).
\end{eqnarray}

Note that the function is independent of the angle $\alpha$ between the dissociation axis and the transition dipole. In other words it is the same for a linear and bent a molecule. 

\bibliography{wiesner1104}
\end{document}